\documentclass[pdflatex,iicol,sn-mathphys-num]{sn-jnl}

\usepackage{graphicx}%
\usepackage{multirow}%
\usepackage{amsmath,amssymb,amsfonts}%
\usepackage{amsthm}%
\usepackage{mathrsfs}%
\usepackage[title]{appendix}%
\usepackage{xcolor}%
\usepackage{chemformula}
\usepackage{textcomp}%
\usepackage{manyfoot}%
\usepackage{booktabs}%
\usepackage{algorithm}%
\usepackage{algorithmicx}%
\usepackage{algpseudocode}%
\usepackage{listings}%
\usepackage{verbatim}  
\usepackage{pdfpages}  

\begin{document}
\title[Article Title]{Data-efficient multi-fidelity training for high-fidelity machine learning interatomic potentials}


\author[1]{\fnm{Jaesun} \sur{Kim}}\email{sunny990912@snu.ac.kr}
\equalcont{These authors contributed equally to this work.}

\author[1]{\fnm{Jisu} \sur{Kim}}\email{kskjs1203@snu.ac.kr}
\equalcont{These authors contributed equally to this work.}

\author[1]{\fnm{Jaehoon} \sur{Kim}}\email{wogns6992@snu.ac.kr}
\author[1]{\fnm{Jiho} \sur{Lee}}\email{dlwlgh6319@snu.ac.kr}
\author[1]{\fnm{Yutack} \sur{Park}}\email{parkyutack@snu.ac.kr}
\author[2]{\fnm{Youngho} \sur{Kang}}\email{youngho84@inu.ac.kr}

\author*[1,3]{\fnm{Seungwu} \sur{Han}}\email{hansw@snu.ac.kr}

\affil[1]{\orgdiv{Department of Materials Science and Engineering}, \orgname{Seoul National University}, \orgaddress{\city{Seoul}, \postcode{08826}, \country{Republic of Korea}}}

\affil[2]{\orgdiv{Department of Materials Science and Engineering}, \orgname{ Incheon National University}, \orgaddress{\city{Incheon}, \postcode{22012}, \country{Republic of Korea}}}

\affil[3]{\orgdiv{AI Center}, \orgname{Korea Institute of Advanced Study}, \orgaddress{\city{Seoul}, \postcode{02455}, \country{Republic of Korea}}}

\abstract{
Machine learning interatomic potentials (MLIPs) are used to estimate potential energy surfaces (PES) from ab initio calculations, providing near quantum-level accuracy with reduced computational costs. However, the high cost of assembling high-fidelity databases hampers the application of MLIPs to systems that require high chemical accuracy. Utilizing an equivariant graph neural network, we present an MLIP framework that trains on multi-fidelity databases simultaneously. This approach enables the accurate learning of high-fidelity PES with minimal high-fidelity data. We test this framework on the \ch{Li6PS5Cl} and In$_x$Ga$_{1-x}$N systems. The computational results indicate that geometric and compositional spaces not covered by the high-fidelity meta-gradient generalized approximation (meta-GGA) database can be effectively inferred from low-fidelity GGA data, thus enhancing accuracy and molecular dynamics stability. We also develop a general-purpose MLIP that utilizes both GGA and meta-GGA data from the Materials Project, significantly enhancing MLIP performance for high-accuracy tasks such as predicting energies above hull for crystals in general. Furthermore, we demonstrate that the present multi-fidelity learning is more effective than transfer learning or $\Delta$-learning an d that it can also be applied to learn higher-fidelity up to the coupled-cluster level. We believe this methodology holds promise for creating highly accurate bespoke or universal MLIPs by effectively expanding the high-fidelity dataset.}

\keywords{Machine learning interatomic potential, Multi-fidelity, Data efficiency, Universal potential}

\maketitle

\section{Introduction}\label{sec1}

Recently, machine learning (ML) has emerged as a powerful tool for addressing various challenges in materials science. In computational materials science, in particular, ML has been effectively used as a surrogate model for ab initio calculations, achieving near-quantum mechanical accuracy with significantly reduced computational costs~\cite{ML_accel_1, ML_accel_2, ML_accel_CSP}. However, creating a high-fidelity training database with accurate ab initio calculations is computationally demanding. For example, band gaps calculated using hybrid functionals align more closely with experimental observations compared to the conventional generalized gradient approximation (GGA) approaches~\cite{HSE_bandgap}. However, the database for band gaps calculated using hybrid functionals is significantly smaller than that using GGA, due to the higher computational costs associated with evaluating exact exchange terms~\cite{MF_bandgap_1, MF_bandgap_2}. Consequently, developing high-fidelity ML models is challenging, as ML approaches typically require large databases for reliable predictions.

One solution to the scarcity of high-fidelity data is to use more abundant but less accurate low-fidelity data. This approach utilizes the observation that low-fidelity data often highly correlates with high-fidelity data, allowing ML models to more efficiently learn the difference between them rather than the entire complex behavior of high-fidelity data from scratch~\cite{Delta_ML_1, Delta_ML_2, Delta_MLIP}. Such multi-fidelity ML strategy has been successfully applied to predict high-fidelity material properties, such as the band gap of inorganic crystals~\cite{MF_bandgap_1, MF_bandgap_2}, electronic levels of molecules,~\cite{MF_HOMOLUMO} and energies of molecular systems~\cite{Delta_ML_1, MF_energy}.

One active application of ML in materials science is the development of machine learning interatomic potentials (MLIPs)~\cite{BPNN, GAP, NequIP, MACE}. MLIPs are models fitted to the energy, atomic forces, and sometimes the stress on a cell, as calculated by ab initio methods. Unlike conventional ML approaches, MLIPs calculate gradient-domain properties (such as force and stress) by taking the derivative of the inferred energy. By enabling large-scale, long-term material simulations at ab initio accuracy, MLIPs significantly expand the boundaries of computational materials research~\cite{AMI_etching, ML_accel_CSP}.

Most of the current MLIPs are trained on ab initio data at the GGA level. Although GGA is sufficient for many applications, some require a higher level of treatment of exchange-correlation energies, such as meta-GGA~\cite{energy_order_1}, the random-phase approximation (RPA)~\cite{RPA_example}, and coupled cluster methods~\cite{CC_example}.  However, preparing a training set for such high-fidelity MLIPs requires significantly more computational time than that required with GGA, because the high-level ab initio methods often have poorer scaling than GGA, and MLIPs often necessitate a large number of configurations to accurately learn the continuous potential energy surfaces (PES) and their derivatives.

In this regard, utilizing multi-fidelity ML approaches for MLIPs is promising because they could enable high-fidelity atomistic simulations with a limited amount of high-fidelity data. Accordingly, several developments have recently emerged in this direction.
For instance, $\Delta$-learning, which uses an ML model to capture differences between low- and high-fidelity data, has been widely studied. Ref.~\cite{delta_review} offers an extensive review on applying $\Delta$-learning to achieve coupled cluster accuracy for PES and force fields for small molecules. For inorganic crystals, ref.~\cite{Delta_MLIP} applied $\Delta$-learning to zirconia, successfully modeling phase transitions at the RPA level. However, the $\Delta$-learning method requires both low- and high-fidelity data for the same configurations (known as a transductive setting~\cite{MF_HOMOLUMO}), which limits its use to datasets that contain different snapshots (referred to as an inductive setting~\cite{MF_HOMOLUMO}). 

Another multi-fidelity approach is to first pretrain MLIPs on low-fidelity data, and then fine-tune them on high-fidelity data using transfer learning. Transfer learning has been applied to achieve coupled cluster accuracy in simulations of small molecules~\cite{TL_mol_1, TL_mol_2}, and at the RPA level for ice crystals~\cite{TL_MLIP_2}. Recently, ref.~\cite{meta_MLIP} employed meta-learning to develop a pretrained MLIP using a large dataset with various ab initio calculation setups, demonstrating improved transferability to target fidelity datasets. While transfer learning can be used in inductive settings, it is susceptible to problems like catastrophic forgetting and negative transfer~\cite{negative_TL, Cata_forget, Cata_negative_TL}.

In this work, we develop a multi-fidelity framework for MLIPs. Unlike transfer learning or $\Delta$-learning, our framework accommodates different levels of fidelity using one-hot encoding, allowing simultaneous training on multiple databases of varying levels of fidelity~\cite{MF_bandgap_1}. We base our implementation on the SevenNet~\cite{SevenNet,NequIP}, an equivariant graph neural network (GNN) MLIP, and modify the model to manage high-fidelity data in the inductive setting, while avoiding issues like catastrophic forgetting commonly associated with transfer learning. 

To test the multi-fidelity training, we employ PBE~\cite{PBE} and strongly constrained and appropriately normed (SCAN)~\cite{SCAN} or regularized-restored SCAN ($\rm{r}^2$SCAN)~\cite{r2scan} functionals as low- and high-fidelity DFT methods, respectively. (An extension to high-fidelity at the coupled cluster levels will be provided in the Discussion section.) The SCAN functionals achieve greater accuracy than PBE due to refined treatment of kinetic energy and dispersion forces and its ability to reduce delocalization errors, resulting in more reliable predictions across diverse materials~\cite{energy_order_2, SCAN_accuracy_2}.

We apply SevenNet-MF in two scenarios.
First, we demonstrate that the present approach can be used to build bespoke high-fidelity MLIPs for specific systems such as $\rm{Li_{6}PS_{5}Cl}$ and $\mathrm{In}_x \mathrm{Ga}_{1-x} \mathrm{N}$ alloys. The computational results indicate that MLIP performance on geometric and chemical environments not directly sampled in the high-fidelity database can be significantly improved through knowledge transferred from the low-fidelity database. 
Second, we train a multi-fidelity pretrained universal MLIPs using the Materials Project~\cite{MaterialsProject} database. Our results indicate that incorporating information from a low-fidelity database can significantly enhance the performance of high-fidelity universal MLIPs.  

\section{Results}\label{sec2}

\subsection{GNN MLIP architecture for multi-fidelity database}\label{sec2subsec1}
\begin{figure*}[!h]
\centering
\includegraphics[width=\textwidth]{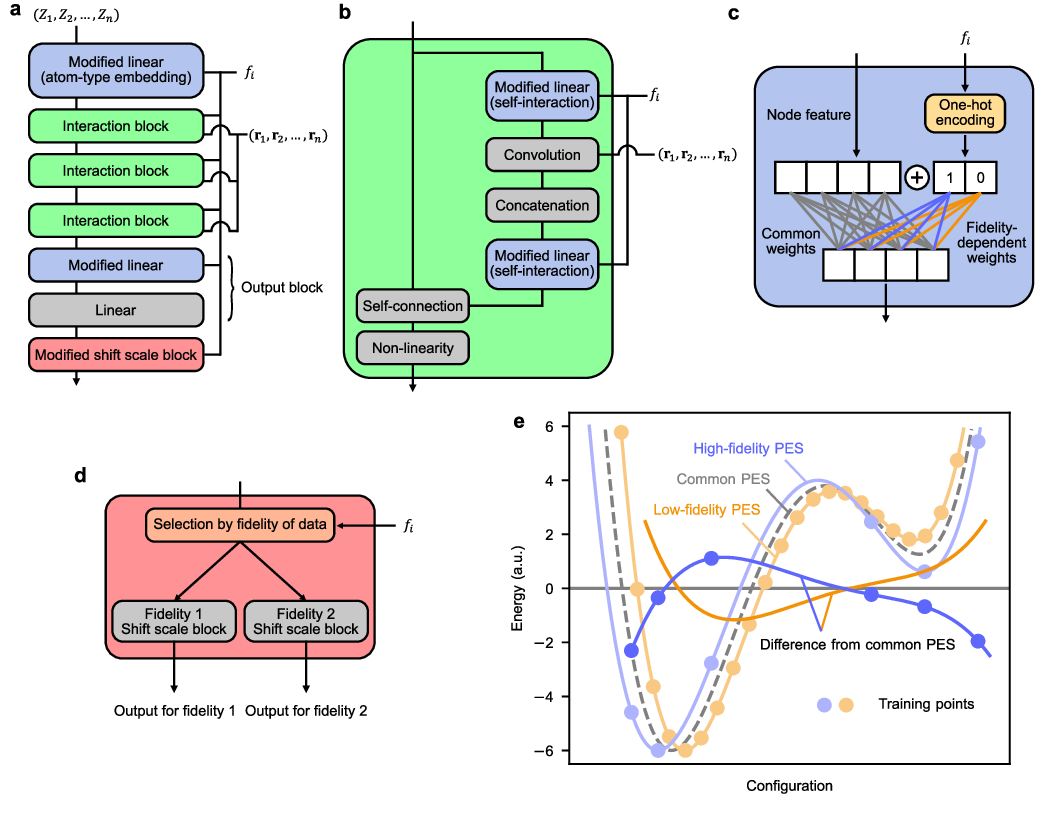}
\caption{\textbf{Architecture of SevenNet-MF.} \textbf{a,} Overall structure of SevenNet-MF with three interaction blocks and two levels of fidelity: $Z_{j}$ and $\mathrm{\mathbf{r}}_{j}$ denote the atomic number and position of the $j$-th atom in the input structure, respectively. $f_i$ represents the fidelity. \textbf{b,} Interaction block. \textbf{c,} Structure of the modified linear layer used in the atom-type embedding, interaction block, and output block. $\oplus$ represents the concatenation of each vector. Weights that are commonly used across the total database are indicated by gray lines, while colored lines represent weights used differently for each fidelity level. The figure shows an example of the data with the first fidelity, so weights in orange lines are unused. \textbf{d,} Structure of the modified shift scale block.  \textbf{e,} Schematic figure showing the PES of the low- and high-fidelity databases. The gray dashed line indicates the common rough trend of the PES observed in both databases, which can be learned using common weights in \textbf{c}. On the other hand, fidelity-dependent weights in \textbf{c} are trained to fit the detailed PES differences between the common PES and the actual ab initio PES.}\label{fig:GNN_arch}
\end{figure*}

We implement a multi-fidelity approach to the equivariant-type GNN architecture of SevenNet~\cite{SevenNet} and name the adapted model SevenNet-MF. For the detailed architecture of the original SevenNet, we refer to refs.~\cite{NequIP, SevenNet}.  We incorporate one-hot encoding of fidelity into the node features used in SevenNet. Figure. \ref{fig:GNN_arch} is a schematic of the model architecture of SevenNet-MF where the unchanged components are shown in gray blocks and the modified parts are highlighted in other colors.
The multi-fidelity framework is implemented within the linear layers (see blue blocks in Fig. \ref{fig:GNN_arch}a--c), which involve trainable matrix multiplication for each node feature.
To be specific, they are the atom-type embedding layers, self-interaction layers in the interaction blocks, and the first self-interaction layer in the output block.

The fidelity of each database is an invariant scalar under rotation and inversion operations of the input structure, i.e., a \texttt{0e} feature according to the notation of the \texttt{e3nn} library~\cite{e3nn}. Thus, we one-hot encode each fidelity and concatenate it to the scalar part of the input node feature of each linear layer (see Fig. \ref{fig:GNN_arch}c).  Fidelity-dependent weights, represented as blue and orange lines in Fig. \ref{fig:GNN_arch}c, are initialized to zero at the start of the training. 
For databases with varying fidelity, statistical parameters like the reference value of total energy can differ depending on the computational settings used, such as the exchange-correlation functional and the choice of pseudopotentials. To reflect this within the model, we modify SevenNet to use different atomic energy shift and scale values for each fidelity database (as shown as red blocks in Fig. \ref{fig:GNN_arch}a,d).

During the inference stage of MLIP, the desired fidelity can be selected by manually specifying a target fidelity as an input to the model. In the present work, unless otherwise specified, we opt for the highest fidelity when using SevenNet trained on a multi-fidelity database.
The present multi-fidelity approach does not alter the convolution pattern of the original SevenNet, allowing us to utilize the parallelization algorithms developed in ref.~\cite{SevenNet} without any changes.
To note, we tested various types of implementation of multi-fidelity and the one in the above yielded robust results (see Methods and Supplementary Section I).

Figure \ref{fig:GNN_arch}e schematically explains the effectiveness of multi-fidelity training. Filled disks represent sampled points in the configuration space. The high-fidelity dataset is sparse, while the abundant low-fidelity database contains diverse configurations. The high- and low-fidelity PES differ but are highly correlated. The multi-fidelity approach uses common parameters for both databases (gray lines in Fig. \ref{fig:GNN_arch}c) to capture the approximate trends of both PESs, or the `common' PES (the gray dashed line in Fig. \ref{fig:GNN_arch}e). Fidelity-dependent parameters are then employed to fit the differences between the ab initio PES and the common PES, enabling smoother interpolation compared to the actual complex PES with a small number of parameters. Thus, even if certain information belongs only to the low-fidelity PBE training set, these details, once learned within common parameters, can be transferred to high-fidelity data through fidelity-dependent parameters.

\subsection{Applications to bespoke MLIP}\label{sec2subsec2}

In this section, we present two specific cases applying SevenNet-MF, which involve \ch{Li6PS5Cl} and $\mathrm{In}_{x}\mathrm{Ga}_{1-x}\mathrm{N}$. 
The argyrodite \ch{Li6PS5Cl} is a popular solid electrolyte, and the accurate computation of Li-ion conductivity using MLIPs is important~\cite{JH_argyrodite}.
On the other hand, the wurtzite $\mathrm{In}_{x}\mathrm{Ga}_{1-x}\mathrm{N}$ quantum wells have a tunable band gap with $x$, making them widely used in light-emitting diodes. To theoretically assess compositional fluctuations under epitaxial growth conditions, it is necessary to estimate relative energies between different alloy configurations~\cite{InGaN_alloy}.
In both applications, it is desirable to use high-fidelity MLIPs at the level of (r$^2$)SCAN to obtain accurate results because of the critical role of lattice parameters and the complex chemical interactions within these materials~\cite{JH_argyrodite, InGaN_alloy}. In the following, we aim to efficiently develop accurate, high-fidelity MLIPs by employing a multi-fidelity training approach, using only a fraction of high-fidelity (r$^2$)SCAN calculations compared to the large low-fidelity training set at the PBE level.

\begin{figure*}[]
\centering
\includegraphics[width=\textwidth]{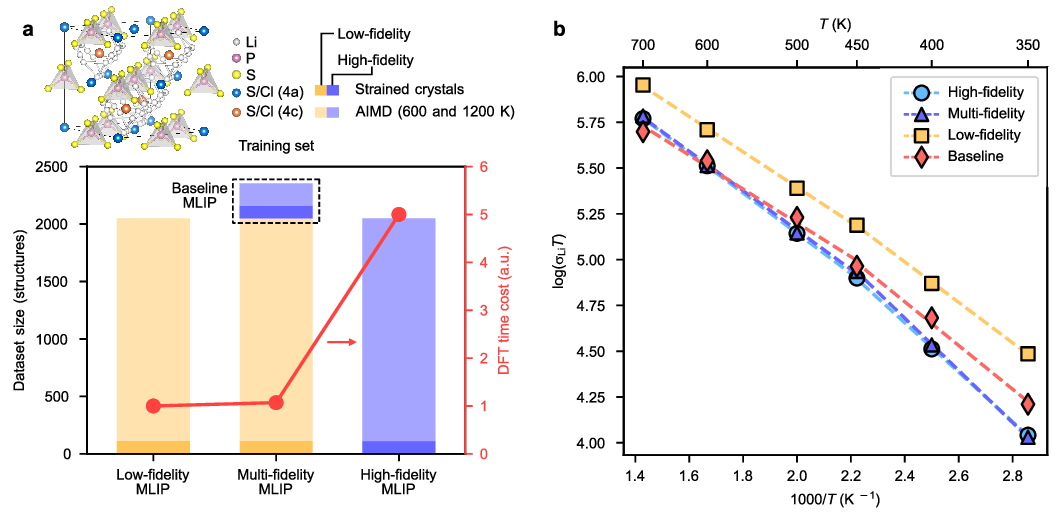}
\caption{\textbf{Multi-fidelity training for \ch{Li6PS5Cl} solid electrolyte.} \textbf{a,} The crystal structure (top left) shows 4a and 4c sites randomly occupied by S or Cl atoms. The training set includes strained crystals and AIMD trajectories, computed with GGA and SCAN functionals. The dataset size for each structure type is displayed at the bottom, alongside the required DFT time, represented as filled disks (right scale). Four types of MLIPs are trained using each dataset. \textbf{b,} Computed Li-ion conductivity ($\sigma_{\rm Li})$ at various temperatures is shown. Dashed lines are linear fits at high temperatures (450--700 K) and low temperatures (350--450 K).
} \label{fig:argyrodite_figure}
\end{figure*}

We first discuss on \ch{Li6PS5Cl}. We consider the maximally disordered system in which S and Cl atoms randomly occupy crystallographic 4a and 4c sites shown at the top of Fig.~\ref{fig:argyrodite_figure}a, which is consistent with experimental analysis~\cite{exp_disorder_1}.  Figure \ref{fig:argyrodite_figure}a also illustrates the construction and size of the training sets for developing MLIPs with various levels of fidelity, as well as the computational costs required for DFT calculations (filled disks). 
The training sets consist of a small set of crystal structures with static deformations and a large set of MD trajectories at elevated temperatures (Methods). The training sets are prepared self-consistently within PBE and SCAN, representing low-fidelity and high-fidelity datasets in Fig.~\ref{fig:argyrodite_figure}a, respectively. The computational costs on the right scale show that high-fidelity datasets are 5 times more expensive than the low-fidelity ones.

The low-fidelity part of the multi-fidelity dataset is the PBE dataset obtained in the above. For high-fidelity part, structures are sampled from the PBE data and one-shot SCAN computations are carried out. While the entire set of static crystals is used, ab initio MD structures are uniformly subsampled by 10\%. The downsampling is significant and the multi-fidelity dataset costs only 10--20\% in addition to the PBE dataset. 
Using the three sets of training data with different levels of fidelity, we train three MLIPs: high-fidelity, multi-fidelity, and low-fidelity. To confirm whether the subsampled high-fidelity dataset is self-sufficient for the accurate MLIP, the baseline MLIP is also developed by training exclusively on the high-fidelity subsampled training set (the dashed square in Fig.~\ref{fig:argyrodite_figure}a). 
Throughout this subsection, the model architecture includes three interaction layers with a maximum degree of spherical harmonics ($l_{\rm{max}}$) of 2 and has 32 channels for each degree of spherical harmonics. The validation errors are all satisfactory (Methods).

With developed MLIPs for \ch{Li6PS5Cl} system, MLIP-MD simulations are conducted from 350 to 700 K and the computed Li-ion conductivity ($\sigma_{\rm Li}$) is shown in Fig. \ref{fig:argyrodite_figure}b (Methods). The $\sigma_{\rm Li}$ at 350 K ($\sigma_{\rm 350 K})$, activation energies at high temperatures (450--700 K) and low temperatures (350--450 K), $E_{\rm a,HT}$ and $E_{\rm a,LT}$ respectively, are provided in Table \ref{tab:Arg_conductivity}. The results with the high-fidelity MLIP agree well with ref.~\cite{JH_argyrodite}. It is found that the multi-fidelity MLIP shows $\sigma_{\rm Li}$ within a 10\% error margin of the high-fidelity model, and notably, $E_{\rm a,LT} - E_{\rm a,HT}$, which is related to the slope change of the linear fit between high and low temperatures, are observed consistently across both high-fidelity and multi-fidelity MLIPs. Conversely, the baseline MLIP exhibits significant deviations in conductivity and underestimates activation barriers. The baseline MLIP also easily results in short Li-Li bonds at temperatures above 600 K, leading to significant energy discrepancies with DFT results (see Supplementary Fig. 5). These results imply that the small size of the high-fidelity sampled SCAN training set is insufficient and the knowledge learned from the low-fidelity PBE training set contributes to increasing the accuracy and stability of MD simulation.

\begin{table}[h]
\caption{\textbf{Results for Li-ion conductivity computed by MLIPs with different levels of fidelity }}\label{tab:Arg_conductivity}%
\begin{tabular}{@{}lccc@{}}
\toprule
Method & $\sigma_{350\rm{K}}$ (mS/cm)  & $E_{\rm a,HT}$ (eV) & $E_{\rm a,LT}$ (eV)\\
\midrule
High-fidelity & 31.4         & 0.218         & 0.268\\
Multi-fidelity       & 30.6 & 0.213     & 0.283\\
Low-fidelity          & 87.4 & 0.191   & 0.219\\
Baseline          & 46.4 & 0.184   & 0.238\\
\botrule
\end{tabular}
\end{table}

\begin{figure*}[]
\centering
\includegraphics[width=\textwidth]{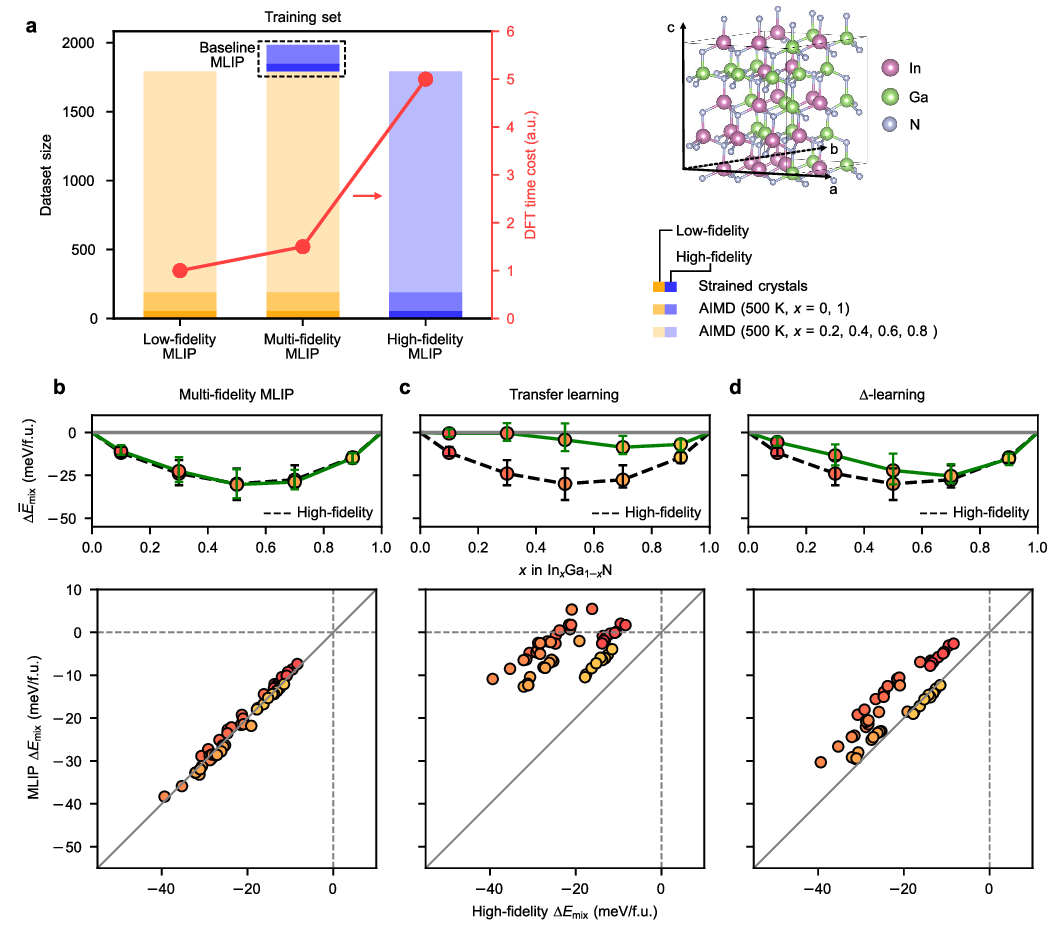}
\caption{\textbf{Multi-fidelity training for In$_x$Ga$_{1-x}$N alloy.} \textbf{a,} Crystal structure (right) and dataset construction for training MLIPs with different levels of fidelity. The required DFT time is shown as filled disks (right scale). The unit cell structure is displayed on the right. The training set consists of crystal structures and AIMD trajectories at 500 K and various alloy compositions. \textbf{b,} The average mixing energy change ($\Delta \bar{E}_{\rm{mix}}$) with respect to the composition $x$, averaged over 10 alloy configurations, is shown. (Top) The dashed line indicates results with the reference high-fidelity MLIP. The error bars indicate the range of mixing energies of the sampled configurations. The parity plot of $\Delta E_{\rm{mix}}$ between multi-fidelity and high-fidelity MLIPs. (Bottom) The color codes represent the composition, consistent with those used in the top figure. \textbf{c,} Results corresponding to transfer learning are displayed. \textbf{d,} Results with $\Delta$-learning are shown.} \label{fig:InGaN_alloy}
\end{figure*}

Next, we discuss on In$_x$Ga$_{1-x}$N. The construction of training set is similar to the case of \ch{Li6PS5Cl} except that among PBE-MD trajectories with $x \in \{0.0,0.2,0.4,0.6,0.8,1.0\}$, only those for pure phases ($x=0,1$) are used for the r$^2$SCAN calculation (Fig.  \ref{fig:InGaN_alloy}a). This is to examine whether SevenNet-MF can interpolate the high-fidelity PES in the composition space, as well as in the configuration space. The computational cost for the multi-fidelity dataset is about 5 times smaller than pure high-fidelity dataset. 
We calculate the mixing energy ($\Delta E_{\rm{mix}}$) defined in the below, which determines the thermodynamic stability of the $\mathrm{In}_{x}\mathrm{Ga}_{1-x}\mathrm{N}$ alloy at 0 K: 
\begin{equation}
\Delta E_{\rm{mix}} = E(\mathrm{In}_{x}\mathrm{Ga}_{1-x}\mathrm{N}) - xE(\rm{InN}) - (1-\textit{x}) \textit{E}(\rm{GaN})\label{eq_emix}
\end{equation}
where $E(\alpha)$ is the energy of $\alpha$ per formula unit (f.u.).

Figure \ref{fig:InGaN_alloy}b shows $\Delta \bar{E}_{\rm{mix}}$ that is averaged over 10 alloy configurations within the 3$\times$3$\times$2 supercell. The $x$ value is selected to be 0.1, 0.3, 0.5, 0.7, and 0.9, which were not sampled in the training set. The lattice vectors in the $ab$ plane are epitaxially fixed to the GaN lattice (top right of Fig. \ref{fig:InGaN_alloy}a). The results for fully relaxed results are shown in the Supplementary Fig. 6. In the top figure of Fig. \ref{fig:InGaN_alloy}b, $\Delta \bar{E}_{\rm{mix}}$ of multi-fidelity MLIP show similar values to those of the high-fidelity MLIP (the dashed line) with a low MAE of 5.51 meV/f.u. The parity plot in the bottom also displays a high correlation with $R^2$ of 0.98, indicating that relative energies among different alloy configurations are predicted with high accuracy. In contrast, the results with baseline MLIP in the Supplementary Fig. 7 exhibit significant errors, which is likely due to the lack of information about In-Ga interactions.

For the example of the $\mathrm{In}_{x}\mathrm{Ga}_{1-x}\mathrm{N}$ alloy, we compare the present multi-fidelity framework with other approaches, namely transfer learning and $\Delta$-learning, which can be also employed to improve the fidelity of MLIPs (Methods). As illustrated in Fig. \ref{fig:InGaN_alloy}c,d, while transfer learning and $\Delta$-learning capture the overall trends in $\Delta E_{\rm{mix}}$, their accuracy is significantly lower compared to the multi-fidelity MLIP, with $R^2$ values of $-4.55$ and $0.31$ in the parity plots, respectively. The decline in transfer learning performance is likely due to the MLIP being overfitted to the binary crystal in the r$^2$SCAN training set, leading to higher errors in predicting the equilibrium energy of the ternary alloy. This issue of catastrophic forgetting could potentially be alleviated by advanced transfer learning techniques such as layer freezing or elastic weight consolidation~\cite{Cata_forget}. On the other hand, the reduced accuracy in $\Delta$-learning may result from the absence of In-Ga interactions in the training data used for learning the difference in PES between low- and high-fidelity ab initio calculations, limiting its ability to capture detailed chemical interactions.

\subsection{High-fidelity pretrained universal MLIP}\label{sec2subsec3}
Recently, pretrained universal MLIPs~\cite{MACE_MP_0, PFP, SevenNet, M3GNet, CHGNet} have gained significant attention as they can bypass the extensive development time typical of bespoke MLIPs, which is due to iterative procedures and trial-and-error in selecting training sets~\cite{STAM}.  Most universal MLIPs are trained on a large crystal dataset computed within the semilocal functional, such as PBE. However, databases like the Materials Project perform r$^2$SCAN calculations on high-priority crystals (stable or metastable crystals in the PBE database) to provide more accurate formation energies and equilibrium volumes. In the same vein, it is worthwhile to develop universal MLIPs that can produce r$^2$SCAN energies, though preparing the training set would take a long time due to the increased computational costs and a huge number of crystals required for training universal MLIPs. 

We address the challenges outlined above within a multi-fidelity framework. To this end, we utilize MPF.2021.2.8 data computed using the PBE functional~\cite{MaterialsProject,M3GNet} as a low-fidelity dataset and r$^2$SCAN data from the Materials Project (hereafter referred to as MP-r$^2$SCAN) as a high-fidelity dataset. The number of crystals included is 60,000 for MPF.2021.2.8 and 32,000 for MP-r$^2$SCAN. Each dataset includes relaxation trajectories, but we note that configurations in MP-r$^2$SCAN are close to equilibrium, exhibiting small forces and stress tensors.

We first examine how the large low-fidelity data influences accuracy and determine how much high-fidelity data is necessary to achieve sufficient accuracy. To this end, we reserve 10\% of the crystals in MP-r$^2$SCAN as a test set. From the remaining 29,000 crystals, we sample 10\%, 20\%, 50\%, and 100\% according to three strategies: random sampling, human intuition, and unsupervised learning. Human intuition involves evenly sampling the chemical space by considering basic chemical and structural information, such as constituent elements and crystal space groups. On the other hand, unsupervised learning involves sampling representative crystals through a clustering approach, described as the DIRECT sampling method in the ref.~\cite{DIRECT}. Detailed descriptions of the sampling methods are provided in the Methods section.

\begin{figure*}[htbp!]
\centering
\includegraphics[width=1\textwidth]{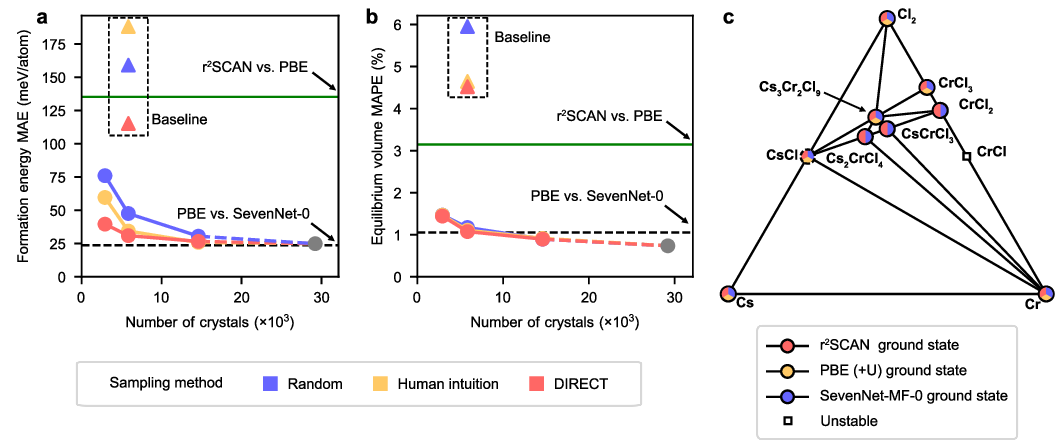}
\caption{\textbf{Multi-fidelity training for pretrained universal MLIP.} \textbf{a,} Mean absolute error (MAE) for the formation energy of the pretrained model with different sizes of the r$^2$SCAN training set, counted by the number of crystals. The MPF.2021.2.8 data are used as a low-fidelity dataset. The leftmost and rightmost models sample 10\% and 100\% of the total training set. Three sampling methods for choosing crystals from the whole training set are shown in different colors. The baseline data are obtained by the single-fidelity model trained solely with the 20\% of the r$^2$SCAN training set, without using low-fidelity PBE data. The horizontal solid line represents the error of the PBE data with respect to the r$^2$SCAN data. The horizontal dashed line indicates the error for SevenNet-0 with respect to PBE, marking the error level of the state-of-the-art pretrained model. \textbf{b,} Corresponding results for mean absolute percentage error (MAPE) of the equilibrium volume. \textbf{c,} Phase diagram for Cs-Cr-Cl constructed using PBE, r$^2$SCAN, and SevenNet-MF-0.  SevenNet-MF-0 is the pretrained multi-fidelity universal MLIP that produces r$^2$SCAN results for a wide range of crystals. The stable phases identified by each method are marked in colors. The dashed circle for CsCl means that the ground-state structure differs between r$^2$SCAN (or PBE) and SevenNet-MF-0. CrCl is found to be unstable for every method.} \label{fig:GP_MLIP_result}
\end{figure*}

We test the performance of the universal multi-fidelity MLIP against the crystals in the test set by fully relaxing the structures within the r$^2$SCAN mode and comparing the formation energy (Fig. \ref{fig:GP_MLIP_result}a) and equilibrium volume (Fig. \ref{fig:GP_MLIP_result}b) with the reference data. The chemical potentials used to compute the formation energies are calculated within the multi-fidelity MLIP. 
Our results show that the mean errors quickly converge to the values obtained using 100\% of the training set, regardless of the sampling methods used, though DIRECT sampling yields the best performance.
For comparison, the PBE values are shown as horizontal lines, confirming that the multi-fidelity MLIPs produce results close to the $\rm{r}^2$SCAN data as intended. The corresponding errors between PBE and SevenNet-0 (a pretrained model trained on the MPtrj dataset~\cite{MaterialsProject, CHGNet}) are shown as horizontal dashed lines, indicating that the current multi-fidelity MLIPs achieve the accuracy of the state-of-the-art pretrained model. Meanwhile, the triangles in Fig. \ref{fig:GP_MLIP_result}a,b represent results from the baseline MLIP trained solely on 20\% of the $\rm{r}^2$SCAN training set, confirming that the large PBE dataset significantly contributes to improving prediction accuracy.  

Next, we develop a pretrained universal MLIP, dubbed SevenNet-MF-0, which can predict energies and equilibrium structures at the r$^2$SCAN level for a wide range of crystals. We employ the complete MP-r$^2$SCAN dataset and a subsampled MPtrj dataset (Methods) as high- and low-fidelity training data, respectively. 
We test the practical application of SevenNet-MF-0 by using it to construct high-fidelity phase diagrams that represent phase stability. Predicting the phase stability of crystals is a challenging task, as the stability of competing phases can be affected by errors of just a few tens of meV/atom.  

Figure \ref{fig:GP_MLIP_result}c compares the phase diagram of Cs-Cr-Cl, constructed using PBE, $\rm{r}^2$SCAN, and SevenNet-MF-0 (Methods). It is found that the stable ground states predicted by PBE and r$^2$SCAN  differ significantly, with PBE predicting high energies above the hull of \ch{CrCl2}, \ch{Cs2CrCl4}, and \ch{CsCrCl3} by 70, 30 and 68 meV/atom, respectively, which are the ground states in r$^2$SCAN. On the other hand, the phase diagram constructed with SevenNet-MF-0 shows significant improvements, correctly predicting the phase stability of all stable compositions in the r$^2$SCAN phase diagram. However, for CsCl, SevenNet-MF-0 predicts the space group of the ground state as Fm$\bar{3}$m rather than Pm$\bar{3}$m, as observed in r$^2$SCAN and experiments~\cite{CsCl_exp}. This discrepancy seems to stem from the low resolution of SevenNet-MF-0 near the convex energy hull, as the energy difference between the two structures is only 7 meV/atom in r$^2$SCAN.  
Such subtle differences could be refined by a small portion of additional r$^2$SCAN calculations, using a mixing scheme as suggested in ref.~\cite{functional_mixing}. Similar results are also confirmed for other phase diagrams of Li-Zr-O, Ag-Te-O, Na-Y-F and Ce-Se-N systems, as provided in Supplementary Section IV.


In spite of the high accuracy demonstrated by SevenNet-MF-0, we note the deficiency of the model in predicting off-equilibrium properties. For instance, the phonon dispersion curves for unary and binary crystals, as shown in Supplementary Fig. 10, display significant softening of optical modes compared to the reference r$^2$SCAN or SevenNet-0 results.
This is consistent with recent reports that general-purpose MLIPs trained on near-equilibrium configurations tend to soften the PES, leading to an underestimation of phonon frequencies~\cite{Phonon_under}. The r$^2$SCAN data provided by the Materials Project currently consists of r$^2$SCAN relaxation trajectories that begin from GGA equilibrium configurations. As a result, MP-r$^2$SCAN exhibits a narrower distribution of forces (and stresses) around zero compared to the PBE database, which contributes to the soft phonon frequencies observed with SevenNet-MF-0 (see Supplementary Section VI). This issue could be addressed by properly augmenting both the low- and high-fidelity databases with high-force configurations.

\section{Discussion}\label{sec3}

Our multi-fidelity GNN approach can also be applied to databases computed using methods beyond (semi)local functionals, such as coupled cluster approaches, which are considered the ``gold standard" in quantum chemistry but scale less favorably, with a complexity worse than $\mathcal{O}(N^5)$ compared to the $\mathcal{O}(N^3)$ scaling of GGA approximations. We apply SevenNet-MF on the ethanol and aspirin molecules using the MD17 dataset~\cite{MD17, sGDML}, which includes calculations at the CCSD(T) and CCSD levels of theory, respectively. Our results show that by using only 10\% of the original coupled cluster database, we can successfully reproduce the PES of coupled cluster methods with significantly enhanced force predictions. A more detailed analysis of the energy and force errors, as well as the PES for ethanol and aspirin, is provided in Supplementary Section VII. 

The present multi-fidelity framework can be extended to non-GNN MLIPs that utilize simpler model architectures, such as the Behler-Parrinello neural network (BPNN)~\cite{BPNN} and Gaussian Approximate Potential (GAP)~\cite{GAP}, by incorporating additional one-hot encoded fidelity information into the input vector. However, employing relatively shallow models like these for multi-fidelity training may be insufficient for predicting properties in the gradient domain. For instance, a simplified BPNN might resemble a linear regression model that fits the PES using symmetry function vectors. In such cases, providing one-hot encoded fidelity information only enables a constant shift in the PES. Consequently, for the same input structure, the model would produce identical force and stress outputs regardless of the given fidelity, failing to adjust gradient-domain properties across different levels of fidelity. In contrast, equivariant GNN architectures enable fidelity-related features to interact directly with geometrical information through deep layers and graph convolution, thereby successfully capturing fidelity-dependent force and stress.

Another potential application of SevenNet-MF is to utilize multiple databases with varying fidelity and configurations. For example, one could train on inorganic crystals from the Materials Project and ab initio molecular databases calculated with hybrid functionals~\cite{Molecule_DB_1, Molecule_DB_2}. By incorporating high-fidelity information from both inorganic and molecular systems, we anticipate that our multi-fidelity approach could be used to develop the ultimate general-purpose MLIPs. It is worth noting that our implementation of SevenNet-MF is capable of handling multiple databases with varying data labels. This feature is particularly useful since molecular databases often lack stress data, whereas the Materials Project includes energy, force, and stress data.

In conclusion, we presented the multi-fidelity MLIP framework, SevenNet-MF, which significantly advances the prediction of high-fidelity PES with limited high-fidelity data. By integrating low- and high-fidelity data through an equivariant graph neural network, this approach not only enhances the accuracy of materials simulations but also optimizes the use of computational resources. We believe this methodology holds promise for creating highly accurate tailored or universal MLIPs by effectively expanding the high-fidelity dataset.

\section{Methods}\label{sec4}

\subsection{Model architecture selection and training}

We test modifications to the model architecture introduced in section \ref{sec2subsec2} to maximize the knowledge transferred from the low-fidelity database. We train a multi-fidelity model on the $\mathrm{In}_{x}\mathrm{Ga}_{1-x}\mathrm{N}$ training set, using binary crystal configurations as high-fidelity databases. We then evaluate the MAE of energy, force, and stress predictions on the $\mathrm{In}_{x}\mathrm{Ga}_{1-x}\mathrm{N}$ alloy configurations calculated with r$^2$SCAN.

First, we test the effects of applying a fidelity-wise shift or scale. Second, we assess whether utilizing a modified linear layer in each block improves generalizability. Our results indicate that using fidelity-wise energy shifts and modified linear layers in interaction blocks and the output block yields robust results (see Supplementary Section I). Notably, the optimized architecture could be dependent on the distribution of training set and target simulations. However, this would be intractable especially for pretrained universal MLIPs utilizing large number of data and parameters. Therefore, we adopt this architecture for training every MLIPs used in this study.

For the SevenNet-MF-0 and baseline models trained on the Materials Project database, we use hyperparameters similar to those in refs.~\cite{SevenNet, GNoME}. The model architecture includes five interaction layers with a maximum degree of spherical harmonics ($l_{\rm{max}}$) of 2. We employ different numbers of channels for each degree of spherical harmonics: 128 channels for \texttt{0e}, 64 channels for \texttt{1e}, and 32 channels for \texttt{2e} features (i.e., \texttt{128$\times$0e+64$\times$1e+32$\times$2e} following the \texttt{e3nn} notation). This setup matches the one used in SevenNet-0 to ensure comparable model capabilities. For bespoke MLIPs, we find that reducing model complexity still yields satisfactory performance. Thus, we use a smaller model with three interaction layers and 32 channels with $l_{\rm{max}}=2$ in $\ch{Li6PS5Cl}$ and $\mathrm{In}_{x}\mathrm{Ga}_{1-x}\mathrm{N}$ examples and 64 channels with $l_{\rm{max}} = 3$ in molecular systems.

For model training, we use the Huber loss function (equation (\ref{eq_Huber})).
\begin{equation}
\mathcal{L}_{\rm{Huber}}(\hat{y}, y, \delta) = \begin{cases}
 \frac{1}{2}(\hat{y} - y)^2,                   & |\hat{y} - y| \le \delta, \\
 \delta\ \cdot \left(|\hat{y} - y| - \frac{1}{2}\delta\right), & \text{otherwise.}\end{cases}\label{eq_Huber}
\end{equation}
The total loss function, incorporating energy, force, and stress loss, is defined by equation (\ref{eq_total_loss}), 
\begin{align}
\mathcal{L}= &\frac{1}{M} \sum_{i=1}^{M} w^{\rm{E}}_{f_{i}}\mathcal{L}_{\mathrm{Huber}}\left(\frac{\hat{E_i}}{N_{i}},\frac{E_i}{N_{i}},\delta\right)\nonumber\\
&+\frac{\lambda_{\rm{F}}}{3M\sum_{i}^{M}N_{i}} \sum_{i=1}^{M}\sum_{j=1}^{N_{i}}\sum_{k}^{3} w^{\rm{F}}_{f_{i}}\mathcal{L}_{\mathrm{Huber}}\left(\hat{F}_{i,j,k},F_{i,j,k},\delta\right)\nonumber\\
&+\frac{\lambda_{\rm{S}}}{6M} \sum_{i=1}^{M}\sum_{l=1}^{6} w^{\rm{S}}_{f_{i}}\mathcal{L}_{\mathrm{Huber}}\left(\hat{S}_{i,l},S_{i,l},\delta\right)\label{eq_total_loss}
\end{align}
where $M$ indicates the number of data in a batch, $E_i, F_{i,j,k}$, and $S_{i,l}$ denotes total energy, $k$-th component of force acting on $j$-th atoms, and $l$-th component of stress (following the Voigt notation) of $i$-th data in batch inferred by MLIP model, respectively. $\hat{E}_i, \hat{F}_{i,j,k}$, and $\hat{S}_{i,l}$ has similar meaning with $E_i, F_{i,j,k}$, and $S_{i,l}$ but indicate the ground truth value calculated with ab initio calculation. When calculating loss, energies, forces, and stresses are in units of eV, eV/\AA, and kbar, respectively. $\lambda_{\rm{F}}$ and $\lambda_{\rm{S}}$ are the weight ratios for force and stress loss, respectively. We set $\lambda_{\rm{F}}$ to 1 and $\lambda_{\rm{S}}$ to 0.01. We also find that assigning additional weight to high-fidelity data enhances the accuracy of the high-fidelity channel in multi-fidelity training. Therefore, we apply specific weights to the energy, force, and stress losses ($w^{\rm{E}}_{f_{i}}$, $w^{\rm{F}}_{f_{i}}$, $w^{\rm{S}}_{f_{i}}$) based on $f_i$, which indicates the fidelity of $i$-th data. For the MP-r$^2$SCAN database, used in constructing the pretrained universal MLIPs, we assign weights of $w^{\rm{E}}_{f_{i}} = w^{\rm{F}}_{f_{i}} = w^{\rm{S}}_{f_{i}} = 7$, while for all other cases, the weights are set to 1. The model is trained using the Adam optimizer~\cite{Adam} with hyperparameter of $\epsilon=10^{-8}$, $\beta_{1}=0.9$ and $\beta_{2}=0.999$.

Notably, after training, the multi-fidelity model achieves satisfactory training and validation errors for both high- and low-fidelity datasets, demonstrating the success of using both databases simultaneously. For example, the \ch{Li6PS5Cl} system achieves root mean squared errors (RMSE) of 1.8 meV/atom, 57.4 meV/\AA, and 0.64 kbar for energy, force, and stress in the validation set, which is comparable to the RMSE reported in ref. \cite{JH_argyrodite}. Similarly, SevenNet-MF-0 achieves mean absolute errors (MAE) of 10.8 meV/atom, 18.3 meV/\AA, and 0.58 kbar for energy, force, and stress in the training set, further confirming the success of multi-fidelity training in developing a pretrained universal MLIP.

\subsection{$\Delta$-learning and transfer learning}
For $\Delta$-learning, we first subtract the energy, force, and stress values calculated with the PBE functional from those in the r$^2$SCAN database consisting of binary crystal configurations. We then train SevenNet on the computed differences to develop the ``$\Delta$ model". The inferred energy, force, and stress from this model are added to the corresponding predictions of a low-fidelity MLIP, which is a SevenNet model trained on the entire low-fidelity database.

For transfer learning, we begin by training SevenNet on the full low-fidelity database. Then, we initialize the trainable parameters of SevenNet using the low-fidelity model and train it on the high-fidelity database with binary crystal configurations. To preserve the knowledge from the pretrained model, we reduce the learning rate to half of the original value used for training the low-fidelity MLIP, a well-known and effective method to retain previous knowledge during training \cite{lr_reduce}.

\subsection{Database curation}

\subsubsection{$\ch{Li6PS5Cl}$ database}
The training set is constructed using 2$\times$2$\times$2 supercells of $\ch{Li6PS5Cl}$ crystal with DFT calculations. The basic composition of the training set is based on ref.~\cite{JH_argyrodite}. The low-fidelity training set includes 0\%, 50\%, and 100\% Cl@4c strained bulk crystals and ab initio MD trajectories at 600 K and 1200 K for the 50\% Cl@4c crystal. (Cl@4c means the percentage of Cl atoms occupying 4c sites randomly. The rest sites are occupied by S atoms.)

All lattice parameters for the $\ch{Li6PS5Cl}$ crystals used in the training set are calculated by the PBE functional. For the strained bulk crystals, a single cell is used for both the 0\% and 100\% Cl@4c crystals. For the 50\% Cl@4c crystal, three distinct cells are employed, each with S/Cl and Li atoms randomly arranged to represent different disorder configurations. For the MD simulations, two distinct cells for the 50\% Cl@4c crystal are generated using the same random arrangement scheme. Independent MD simulations are conducted at 600 K and 1200 K using both cells, to sample various disordered configurations.

The strained bulk crystals include structures deformed by either shear strain or hydrostatic strain and the deformations are within 5\% of the lattice parameters. Ab initio MD simulations are conducted using the NVT ensemble with a Nos\`e--Hoover thermostat~\cite{Nose_Hoover} and a 2 fs timestep for a total of 10 ps. For the low-fidelity training set, structures are sampled at 20 fs intervals after the initial 300 fs. The high-fidelity training set involves single-point calculations for all structures in the strained bulk crystals and a 200 fs sampling interval for the ab initio MD structures, using only 10\% of the MD configurations from the low-fidelity training set. For reference high-fidelity MLIP, we conduct similar process with low-fidelity training set, except we employ SCAN functional instead of PBE.

DFT calculations are performed using Vienna Ab initio Simulation Package (\texttt{VASP})~\cite{VASP} with PAW PBE pseudopotentials~\cite{PAW_1} and spin-unpolarized calculations. A plane-wave basis set cutoff energy of 300 eV and a {\bf k}-point grid of 2$\times$2$\times$2 per unit cell are used.

\subsubsection{$\mathrm{In}_{x}\mathrm{Ga}_{1-x}\mathrm{N}$ database}
The PBE training set consists of InN and GaN wurtzite crystals under hydrostatic strain within 5\%, and MD simulations of In$_x$Ga$_{1-x}$N with $x \in \{0.0,0.2,0.4,0.6,0.8,1.0\}$ using 3$\times$3$\times$2 and 5$\times$5$\times$1 supercells (the multiplicity is with respect to the conventional cell with in the order of $a$, $b$, and $c$ axes). For alloys, we choose two sets of random distributions of In/Ga for each supercell 
to sample diverse chemical environments. 

The MD simulations are conducted using the NVT ensemble with a Nos\'e--Hoover thermostat at 500 K, a 2 fs timestep, and a total duration of 5 ps.
The lattice vectors for each $x$ are obtained by fully relaxing the supercell volume and shape at 0 K. For $x>0$, another lattice vectors are also used by relaxing the supercell only along the $c$-axis while the $a$- and $b$-axes are fixed to those of GaN, emulating the epitaxially grown structure of $\mathrm{In}_{x}\mathrm{Ga}_{1-x}\mathrm{N}$ in multi-quantum well structures~\cite{InGaN_epitaxy}. 
The structures are sampled for training with the 80 fs interval after the initial 1 ps. 

The high-fidelity training set for multi-fidelity training is obtained by $\rm{r}^2$SCAN single-point calculations on binary crystal configurations. For high-fidelity training set used for reference high-fidelity MLIP, we repeat the same procedure of curating low-fidelity training set, using r$^2$SCAN functional instead of PBE.

DFT calculations are performed using the \texttt{VASP} with projector-augmented-wave (PAW) PBE pseudopotentials and spin-unpolarized calculations. We use a plane-wave basis set cutoff energy of 500 eV, and the {\bf k}-point grid of 5$\times$5$\times$3 for the unit cell, 2$\times$2$\times$2 for the 3$\times$3$\times$2 supercell, and 1$\times$1$\times$3 for the 5$\times$5$\times$1 supercell. In order to accelerate the sampling configurations from ab initio MD simulations, we utilize $\Gamma$-point calculation and then single-point calculate sampled structures with corresponding {\bf k}-point of the supercell.

\subsubsection{Materials Project database}
The Materials Project generates a database using $\rm{r}^2$SCAN functional by performing geometric relaxation on given crystal structures. In February 2024, we retrieved all available structure relaxation task types using the $\rm{r}^2$SCAN functional. Some relaxation trajectories correspond to the same material identification number in the Materials Project. In these cases, to eliminate calculations on supercell structures, we first selected crystals with the minimum number of atoms in the cell. After this procedure, we finally selected the structural relaxation trajectory that was calculated first. For each relaxation trajectory, we selected the initial, middle, and final configurations.

We subsample the high-fidelity training set to see the convergence behavior of MAE in formation energy and equilibrium volume. When subsampling, we assume the scenario of selecting the crystals to undergo r$^2$SCAN relaxation. Therefore, we choose trajectory by trajectory. For instance, when we apply 10\% random sampling, we choose 10\% relaxation trajectories, instead of selecting 10\% configurations of entire MP-r$^2$SCAN database. The sampling method driven by human intuition involves constructing a subsampled set where, for all elements, the binary interactions between elements, as well as the prototype defined using space groups and composition systems, exceed a specific threshold count. This allows the subsampled set to be constructed by formulating a linear problem based on the given conditions. By adjusting the threshold count, the total number of structures can be controlled. We also utilize DIRECT sampling introduced in ref.~\cite{DIRECT} to subsample MP-r$^2$SCAN database. We utilize M3GNet encoder to featurize each initial structure of relaxation trajectory, then utilize DIRECT sampling with those feature vectors. We sample five configurations in each cluster (i.e., $k = 5$), and for the number of clusters, we use 613, 1255, and 3417 for 10\%, 20\%, and 50\% subsampling, respectively. Those numbers are empirically selected to adjust sampling ratio.

For low-fidelity PBE database for pretrained universal MLIP, we utilize MPF.2021.2.8 and MPtrj databases. We observe that utilizing every MPtrj data for multi-fidelity training computationally demanding, thus we subsample configurations with the following rules. First, we obtain configurations that belong to the PBE relaxation trajectory with two or more ionic steps in MPtrj database. Next, we sample the first and the last configurations of the relaxation trajectories.

\subsection{Calculation details of materials simulations}

\subsubsection{$\rm{Li_{6}PS_{5}Cl}$ MD simulations with MLIPs}

First, we utilize MLIP-MD simulations of $\rm{Li_{6}PS_{5}Cl}$ in its unit cell to validate MLIPs, as the unit cell is small enough to be calculated with DFT. This allows us to check whether the MLIPs follow the trends observed in DFT. For MD simulations with MLIPs, NVT simulations are conducted using \texttt{LAMMPS}~\cite{LAMMPS} package. For DFT calculations, we use the same ab initio settings as in the high-fidelity training set.

Next, we calculate the Li-ion conductivity of $\rm{Li_{6}PS_{5}Cl}$ using a 3$\times$3$\times$3 supercell, which provides converged conductivity within an error margin of 10\% as reported in ref.~\cite{JH_argyrodite}. We calculate conductivity at 350, 400, 450, 500, 600, and 700 K. First, a 100 ps NVT MLIP-MD simulation at each temperature is conducted in \texttt{LAMMPS} to equilibrate the structures. Next, each structure is relaxed using Atomic Simulation Environment (\texttt{ASE}) package~\cite{ASE}, maintaining the cubic structure. Finally, a 2 ns NVT MLIP-MD simulation is conducted in \texttt{LAMMPS}, and the mean squared displacement (MSD) of Li ions is calculated for high temperatures (450--700 K). For low temperatures (350 and 400 K), we extend simulation time to 10 ns to ensure convergence of diffusivity. The MSD is linearly fitted over time, excluding the initial non-linear ballistic region, and its slope is used to calculate Li-ion diffusivity ($D_{\rm Li}$) using equation (\ref{eq_diffusivity}). 
\begin{equation}
D_{\rm Li}=\lim_{\textit{t} \to \infty} \frac{\rm{MSD}(\textit{t})}{6\textit{t}}\label{eq_diffusivity}
\end{equation}
Then, we convert Li-ion conductivity ($\sigma_{\rm{Li}}$) from diffusivity using Nernst--Einstein equation:
\begin{equation}
\sigma_{\rm{Li}} = \frac{N_{\rm{Li}}e^2}{Vk_{\rm{B}}T}D_{\rm Li}\label{eq_einstein}
\end{equation}
where $N_{\rm{Li}}, e, V, k_{\rm{B}}, T$ indicates the number of Li ions in the supercell, elementary charge, volume of the simulation cell, Boltzmann constant, and simulation temperature, respectively.
For activation energy, we use the slope of the linear fit of conductivity times temperature versus the inverse of temperature using equation (\ref{eq_activation})
\begin{equation}
\log({\sigma_{\rm{Li}}T}) = -\frac{E_{\rm{a}}}{k_{\rm{B}}}\frac{1}{T}+C\label{eq_activation}
\end{equation}
where $C$ denotes a constant.
For all NVT simulations, we use a Nos\`e--Hoover thermostat with a 2 fs timestep.

\subsubsection{Phase diagram construction}
To construct the phase diagram, we first retrieve all possible crystal structures within the target chemical system from the Materials Project. Hull energy calculations are performed using the \texttt{PhaseDiagram} module from the \texttt{pymatgen} package~\cite{pymatgen}. The phase diagram is then plotted using the calculated hull energies with the \texttt{PDPlotter} module, also provided in \texttt{pymatgen}. For the PBE (+U) and r$^2$SCAN phase diagrams, we use energies from the Materials Project (with corrected energies for PBE (+U)~\cite{MP_correction}). We also relax each phase with SevenNet-MF-0 to generate an MLIP-based phase diagram. For the Cs-Cr-Cl system, we append any missing phases from the r$^2$SCAN database by performing additional r$^2$SCAN calculations using the same settings with Materials Project, as those generated by the \texttt{MPScanRelax} module in \texttt{pymatgen}.

\subsubsection{Phonon calculation}
For phonon calculations, we primarily utilize the \texttt{phonopy} package~\cite{phonopy}. Initially, we relax the geometry of the crystals to find the equilibrium structure using the \texttt{ASE} package combined with SevenNet. For the equilibrium structures, we replicate the cell to create a supercell with at least 5 times the original lattice parameter. We then use \texttt{phonopy} to introduce displacements to the supercell with a magnitude of 0.01 \AA. Finally, we sample high-symmetry {\bf q}-points to plot phonon band structure.

\section{Data availability}
Training data for the MLIPs will be available upon request, except for the open database.

\section{Code availability}
The SevenNet-MF package and the SevenNet-MF-0 model are available at https://github.com/MDIL-SNU/SevenNet.

\bibliography{ms}

\section{Acknowledgements}
This research was supported by the National Research Foundation of Korea (NRF) grant funded by the Korea government (MSIT) (RS-2023-00247245) and the Nano \& Material Technology Development Program through the National Research Foundation of Korea (NRF) funded by Ministry of Science and ICT (RS-2024-00407995).
The computations were carried out at Korea Institute of Science and Technology Information (KISTI) National Supercomputing Center (KSC-2023-CRE-0458) and at the Center for Advanced Computations (CAC) at Korea Institute for Advanced Study (KIAS).
\section{Author contributions}
J.K.$^1$, J.K.$^2$ and S.W. conceived the initial idea. J.K.$^1$ developed SevenNet-MF and Y.P. checked the model structure. J.K.$^1$ and J.K.$^2$ trained and tested models. J.K.$^3$ and J.L. prepared training data for bespoke models. Y.K. and S.H. offered insight and guidance throughout the project. J.K.$^1$, J.K.$^2$ and S.W. prepared the paper. All authors contributed to discussions and approved the paper. (J.K.$^1$: Jaesun Kim, J.K.$^2$: Jisu Kim, J.K.$^3$: Jaehoon Kim)

\section{Competing interests}
The authors declare no competing financial or non-financial interests.

\end{document}